\documentclass[11pt]{article}

\usepackage[T1]{fontenc}
\usepackage{lmodern}

\usepackage[margin=1.1in]{geometry}
\usepackage{booktabs}
\usepackage{array}
\usepackage{graphicx}
\graphicspath{{./}{figures/}}
\usepackage{amsmath,amssymb}
\usepackage[hidelinks]{hyperref}
\usepackage{xurl}
\usepackage{natbib}
\usepackage{caption}
\usepackage{setspace}
\usepackage{microtype}

\title{\textbf{Skill, Luck, or Imitation?\\What Actually Pays in Creator Markets}}
\author{Samiha Tariq\\[0.5em]
\normalsize School of Analytics, Finance and Economics\\
\normalsize Southern Illinois University Carbondale}
\date{\today}

\begin{document}
\maketitle

\begin{abstract}
\noindent
A small share of online creators earn fortunes while most earn almost nothing.
What separates them: skill, luck, or copying what works? This paper measures what
YouTube vloggers actually produce: 3,728 thumbnails, 385 hours of speech (2.9
million words), and video titles and tags, linked to complete view histories of
56,419 videos. Its central sample follows channels from their first upload,
including the many that never took off, which most creator statistics miss. The
data show that what goes into a video does not predict its success: 36
measures of images, speech and text explain at most 1.1\% of the differences in
views within a channel, even though they can tell what kind of video it is.
Practice does not help either: forty more videos without a hit add nothing.
Nothing accumulates except a hit. A channel's first hit raises the views of
everything it publishes afterwards by \textbf{164\%} on average, a result that
survives all 17 robustness checks. Creators respond by copying their hit, both
its look and its topic, yet only copying the topic pays; copying the look adds
nothing measurable. A simple model explains why. A hit is a lasting prize, so
entering can make sense despite long odds, and because the platform's statistics
describe a hit without separating its causes, copying everything about it is the
sensible response. Platforms could redirect creators' effort toward what pays by
showing whether a hit's topic or its look drove its success.

\vspace{0.6em}

\end{abstract}

\clearpage

\section{Introduction}
\label{sec:intro}

A small number of YouTube creators earn millions of dollars. Most earn almost
nothing. The median channel in the data for this paper, followed from its very
first upload, generated \$1.61 in long-form advertising revenue over roughly
four months, for about 185 hours of work. Creative markets have always looked like this. What has never been
settled is why.

Two classic explanations compete. \citet{rosen1981} shows that small
differences in talent can turn into very large differences in earnings when
one performer can serve a huge audience at almost no extra cost. On this view
the stars are genuinely better. \citet{adler1985} shows that stardom can arise
among equally talented performers, because audiences gain more from performers
whom others already follow. On this view the winner is largely an accident of
history.

Both theories predict the same unequal outcomes, so inequality alone cannot
tell them apart. What would tell them apart is quality, and quality cannot be
observed. Studies that measure quality by success assume the answer, and expert
ratings are shaped by knowledge of who succeeded.

This paper takes a different route. Rather than measure quality, it measures
the content itself, in three modalities and for more than a thousand videos. From thumbnails it
records the faces, on-screen text, colour and composition of each image. From
speech it records how fast creators talk, how they open their videos and which
themes they discuss. From titles and metadata it records what each video is
about and how it is labelled. These are the choices that industry advice treats
as the keys to success. They are observable before any outcome occurs, and
creators control them directly. The paper asks whether these characteristics
predict success, and whether creators behave as if they do. The answers turn
out to differ.

The analysis rests on an original multimodal dataset. It
covers 3,728 thumbnails analysed with optical character recognition and deep
learning face detection, 384.7 hours of speech containing 2,940,197 transcribed
words, the titles and tags of the videos analysed, and complete upload histories
covering 56,419 videos from 420 channels. Channels are also sampled by the date they started rather than by
whether they became visible. Nearly every published figure on creator earnings
comes from channels a researcher was able to find, which on a platform where
failure means invisibility is a sample of survivors. In the birth-date sample
used here, half of all new channels (51.4\%) never reach 1,000 lifetime views.

A simple model of demand and creator learning organises
the analysis. Views depend on a channel's baseline level, on experience, on an
audience that grows only through hits, and on the returns to a video's topic
and design. Creators do not know these returns and must learn them from
outcomes. The model yields five testable predictions that separate a
skill-based account of success from one driven by luck and path dependence.

The first finding is that content does not predict success. Across 36 visual,
spoken and textual measures, content explains at most 1.1\% of the variation in
views within a channel, and a flexible machine-learning model does no better
than a linear one. Yet the same measures identify a video's sub-genre with an
AUC of 0.782, so the measurement is sound and the content simply does not
predict returns.

Nothing accumulates except a hit. A channel's first breakout, a video with at
least five times the channel's usual views, raises the views of every video that
follows by 164\% on average. The effect is absent in a placebo on channels that
never broke out, holds across all 17 alternative specifications, and persists across
every later video observed. Experience, by contrast, adds nothing. Forty more
videos without a hit leave performance unchanged. In a setting with constant
practice, instant feedback and strong incentives, there is no learning by doing
\citep{arrow1962}.

Creators respond to a hit by copying it. After a breakout, their next videos
move closer to the winner in two independent ways: in how the thumbnail looks
and in what the video is about. The two measures share no data, yet they move
together. The shift is a sharp jump at the breakout that fades over the
following videos, and it is aimed at the winner specifically.

Only one kind of copying pays. Matching the winner's topic raises views by about
16\% per standard deviation. Matching its look has no measurable effect.
Creators copy both, and only one earns a return.

The paper makes four contributions. First, it builds
a multimodal dataset that measures creative content across images, speech and
text, and it validates that measurement against a known benchmark, which turns
a finding of no effect into evidence that there is no effect to find. Second,
it introduces a sampling frame built on channel birth dates, which describes the
prospects facing someone deciding whether to start rather than the success of
those who already made it. Third, it provides large-scale evidence on how
creators learn from their own hits, separating the part of that learning that
pays from the part that does not. Fourth, it offers an economic account that
connects the findings. Because a hit is lasting, entry can be rational despite a
median return of \$1.61. Because a hit is a single event whose causes cannot be
separated with the statistics and tests creators have, copying every feature of
it is the sensible response. The effort spent copying the look of a hit is created by the
information environment, which makes it a problem that platform design can
solve.

Section~\ref{sec:lit} reviews the related literature and
Section~\ref{sec:theory} presents the theoretical framework.
Section~\ref{sec:data} describes the data and methods. Section~\ref{sec:results}
reports the results and Section~\ref{sec:robust} tests their robustness.
Section~\ref{sec:disc} discusses identification and the economic
interpretation, and Section~\ref{sec:conc} concludes.

\section{Related literature}
\label{sec:lit}

\subsection{What explains superstar outcomes?}

Extreme concentration of rewards is a defining feature of creative markets, and
the literature offers explanations that agree on the shape of inequality but
disagree about its source. In the talent view, small differences in ability are
magnified because one performer can serve an entire market and consumers do not
treat several weaker performers as a substitute for one excellent one
\citep{rosen1981}. In the popularity view, consumers enjoy a performer more when
they know more about the performer's work, and they learn most cheaply by
discussing performers whom others already follow, so audiences converge on
whoever is already popular and stars can emerge among equally talented
performers \citep{adler1985}. A third account links the two through learning.
When early performance is informative about later performance, only those who
succeed early continue, and entrants accept low early earnings for the chance of
later success \citep{macdonald1988}. Because talent is hard to identify in
advance, the later supply of highly talented artists depends on how many young
artists are able to begin careers \citep{alcala2010}.

Empirical work has struggled to separate these accounts because each one fits
the shape of the earnings distribution. Studies of the concert market document
sharply rising prices and a growing share of revenue for the top performers
\citep{krueger2005}, and the organisation of creative industries is shaped by
the fact that demand for a new product is highly uncertain, even for
experienced professionals \citep{caves2000}. That uncertainty makes wider entry
valuable. When the appeal of a new work cannot be predicted at the time it is
made, letting more works reach the market produces more successes: once authors
could bypass publishers, self-published books grew by 2013 to about a tenth of
bestseller listings \citep{waldfogel2015}. Changes in supply alone can produce
the same shape. In a model of how digitisation affects media industries, a fall in the fixed cost of basic products is enough to generate
superstars and a long tail at the same time \citep{weeds2012}. Recent work
extends superstar theory to online creators, concluding that the classic superstar mechanisms
still apply to online influencers \citep{gaenssle2021} and examining them in a
sample of 200 established YouTube stars \citep{budzinski2018}. Samples of
established stars, however, condition on the outcome of interest. The present
paper addresses both limitations at once. It measures content directly, so the
talent view can be tested through the characteristics said to carry quality,
and it samples channels at birth, so the analysis includes the creators who
never became stars.

\subsection{Luck, social influence and path dependence}

If winners are not better, something else must select them. Theories of
increasing returns show how small historical accidents can lock in outcomes, so
that which option wins depends on the order of events rather than on merit
\citep{david1985,arthur1989}. Theories of social learning show how people can
rationally follow the choices of others instead of their own information,
producing herds that are fragile and often arbitrary
\citep{banerjee1992,bikhchandani1992}. Experiments confirm that these forces
operate in cultural markets. When listeners can see what others have
downloaded, identical songs reach very different levels of success in parallel
markets \citep{salganik2006}, displayed popularity partly fulfils itself
\citep{salganik2008}, and a single early positive vote raises the final ratings
of online comments \citep{muchnik2013}. Field evidence points the same way:
appearing on a bestseller list modestly raises a book's later sales
\citep{sorensen2007}. Producers follow popularity as well. In markets for news,
both novel stories and stories on topics already popular with readers are
associated with higher sales, the pull of a popular story fades within one to
two weeks, and competing outlets often converge on the same story
\citep{ho2015}.

What this literature has not measured is how long the advantage from a
producer's own success lasts, or what the producer does in response. Experimental designs randomise exposure to fixed products, so they
cannot follow how a producer's later output changes. The present paper
measures the lasting effect of a first breakout on every later video and uses
the multimodal data to track how the creator's own content changes in response.

\subsection{Returns to entry and learning from noisy feedback}

Persistent entry into occupations with poor average returns is well
documented. Most self-employed workers earn less than they would in paid
employment \citep{hamilton2000}, and entrepreneurs earn no higher returns on
their private businesses than on public equity despite bearing far more risk
\citep{moskowitz2002}. Creative markets show the same pattern. Piracy cut music
revenue without reducing the entry of new artists, which \citet{gans2015}
explains through artists who are time inconsistent in how they weigh fame
against fortune. Other proposed explanations include non-financial benefits,
overconfidence \citep{camerer1999}, a range of behavioural factors
\citep{astebro2014}, and selection on early success when outcomes reveal ability
\citep{jovanovic1982,macdonald1988}. Yet money matters to creators at the
margin. When YouTube creators lost subscription income for reasons outside
their control, they published fewer videos and shifted toward sponsored content
and affiliate marketing \citep{panjwani2026}. A persistent prize offers an
explanation of both facts that requires neither mistaken beliefs nor
non-financial rewards: creators respond to money, and entry can still be rational
because what it buys is a chance at a lasting prize, not the average return.
The evidence on hits in this paper speaks to it directly.

The learning side of the problem is equally unsettled. The benchmark expectation
is that productivity rises with experience \citep{arrow1962}. When outcomes are
noisy, however, organisations can credit success to features that did not cause
it \citep{levitt1988}, learning from visible successes while failures go unseen
leads to wrong conclusions \citep{denrell2003}, and the choice between repeating
what worked and trying something new becomes hard \citep{march1991}. These
arguments are rarely tested on individual producers, because a test requires
measuring what a producer changes after a success and whether those changes pay.
By measuring content along separate visual and topical dimensions and estimating
the return to each, this paper shows which part of creators' learning is
mistaken and which part is correct.

\subsection{Images, speech and text as economic data}

A growing literature treats unstructured content as economic data. Methods for analysing text are increasingly used in economics \citep{gentzkow2019}, and machine
learning is increasingly used to construct variables that then enter
conventional econometric models \citep{mullainathan2017}. Interpretable image
features have been shown to predict demand for rental listings
\citep{zhang2022} and engagement with social media posts \citep{li2020}. This
paper applies these tools to a creative market and combines three modalities,
images, speech and text, in a single design. It uses machine learning for
measurement and ordinary regression for inference, and it keeps every feature
interpretable, so that an estimated effect can be read as the effect of a
specific production choice.

\section{Theoretical framework}
\label{sec:theory}

This section presents a simple framework that organises the empirical analysis.
It nests the skill-based and luck-based accounts of creative success as special
cases and generates predictions that the data can test.

\subsection{Demand and the value of a hit}

A creator $i$ publishes videos in sequence $s = 1, 2, \ldots$ Each video has a
topic $x^{T}_{is}$ and a visual design $x^{V}_{is}$, both measured relative to
the creator's usual content. Views follow
\begin{equation}
\log_{10} v_{is} \;=\; \alpha_i \;+\; \lambda\, s \;+\; A_{is}
\;+\; \theta_T\, x^{T}_{is} \;+\; \theta_V\, x^{V}_{is} \;+\; \varepsilon_{is},
\label{eq:model}
\end{equation}
where $\alpha_i$ is the channel's baseline level, $\lambda$ is the return to
experience, $A_{is}$ is the audience the channel has accumulated from past hits,
$\theta_T$ and $\theta_V$ are the returns to topic and design, and
$\varepsilon_{is}$ is a demand shock with a heavy right tail.

A breakout is a video whose shock exceeds a high threshold $\kappa$. A breakout
brings new viewers, some of whom return, and it can raise the channel's standing
with the recommendation system. The audience therefore grows only through hits:
\begin{equation}
A_{i,s+1} \;=\; A_{is} \;+\; \delta\, \mathbf{1}[\varepsilon_{is} > \kappa],
\label{eq:audience}
\end{equation}
so a first breakout raises the level of every later video by $\delta$.

The two classic accounts correspond to different parameter values. In a
skill-based account, quality shows up in content and improves with practice, so
the returns $\theta$ are large relative to the spread of $\varepsilon$ and
$\lambda > 0$. In an account driven by luck and path dependence, content carries
little information about demand, experience adds nothing ($\lambda = 0$), and a
channel's position changes only through realised hits whose effect persists
($\delta > 0$).

\subsection{Learning from a single hit}

The creator does not know $\theta = (\theta_T, \theta_V)$ and learns it from
outcomes. Suppose the creator holds independent normal priors
$\theta_j \sim N(0, \tau^2)$ and, for tractability, treats demand shocks as
normal with variance $\sigma^2$. After a breakout on a video with
characteristics $x^{*} = (x^{*}_{T}, x^{*}_{V})$, the creator observes a single
outcome $y^{*} = \theta_T x^{*}_{T} + \theta_V x^{*}_{V} + \varepsilon^{*}$,
measured net of the channel's level. Bayes' rule gives the posterior mean
\begin{equation}
E\left[\theta_j \mid y^{*}\right] \;=\;
\frac{\tau^2\, x^{*}_{j}}{\tau^2\left(x^{*2}_{T} + x^{*2}_{V}\right) + \sigma^2}\; y^{*},
\qquad j \in \{T, V\}.
\label{eq:posterior}
\end{equation}

Equation~\eqref{eq:posterior} contains the central point. The update for each
characteristic depends on how distinctive that characteristic was in the
winning video, $x^{*}_{j}$, and on the size of the success, $y^{*}$. It does not
depend on the true return $\theta_j$. A single outcome cannot reveal which
characteristic produced it, so the creator raises the estimated return to every
distinctive feature of the hit at once. The same formula shows why ordinary
videos teach little: when shocks are large relative to the returns, so that
$\sigma^2$ is large relative to $\tau^2$, each moderate outcome moves beliefs
only slightly.

Suppose the creator then chooses the next video's characteristics $x$ to
maximise expected views net of a quadratic cost of changing style,
$E[\theta \mid y^{*}]'x - (c/2)\lVert x \rVert^2$. The optimal choice is
$x_j = E[\theta_j \mid y^{*}]/c$, which is proportional to $x^{*}_{j}$. The next
video therefore moves toward the winner in every dimension, whether or not that
dimension carries demand. As later outcomes arrive, the weight on the single
extreme success falls. Because $y^{*}$ contains a large shock, the first update
overshoots and later evidence pulls it back, so imitation jumps at the breakout
and then fades.

Whether imitation pays depends only on the true returns. Moving toward the
winner by $x_j$ changes expected log views by $\theta_j x_j$. If design carries
no demand ($\theta_V = 0$) while topic does ($\theta_T > 0$), then topical
imitation raises views and visual imitation does not, even though the creator
imitates both.

\subsection{Entry}

Consider a new creator who plans to publish $S$ videos, each of which breaks out
with probability $p$. Let $r$ be the expected revenue of a video before any hit
and $m = 10^{\delta}$ the multiplier that a hit applies to every later video.
Counting only the first hit, which gives a lower bound on the value of the
prize, the probability that the first hit has occurred before video $s$ is
$1 - (1-p)^{s-1}$, and expected revenue over the creator's horizon is
\begin{equation}
R(S) \;=\; r \sum_{s=1}^{S}\Big[1 + (m-1)\big(1 - (1-p)^{s-1}\big)\Big]
\;=\; rS \;+\; r(m-1)\left[S - \frac{1-(1-p)^{S}}{p}\right].
\label{eq:entry}
\end{equation}

The first term is what the creator earns without a hit. The second is the value
of the lasting prize, and it grows with the horizon $S$. For a creator who
expects to publish for a long time, the prize can dominate the value of entry
even when $p$ is small and the typical early outcome is close to zero. Entry into
a market with a very low median return can therefore be rational. Related logic
appears in models where early success reveals ability \citep{macdonald1988}. In
this framework the prize persists because the audience persists, whether or not
the hit reveals anything about ability.

\subsection{Testable predictions}

The framework yields five predictions that separate the two accounts.

\begin{description}
\item[P1. Content predicts little.] Within a channel, content characteristics
explain little of the variation in views. Under a skill-based account they would
explain a substantial share. Content can still distinguish genres, because
genres differ in content even when content does not drive demand, and this
provides a check on the measurement.

\item[P2. A hit is lasting.] A first breakout raises the level of every later
video ($\delta > 0$).

\item[P3. Experience adds nothing.] Once hits are accounted for, views do not
rise with a channel's number of videos ($\lambda = 0$).

\item[P4. Creators imitate their hits.] After a breakout, creators move toward
the winning video in every observable dimension. The movement is directed at the
winner rather than at other past videos, with a sharp change at the breakout
that later fades.

\item[P5. Imitation pays only where demand responds.] Similarity to the winner
raises views only along dimensions with $\theta_j \neq 0$.
\end{description}

Section~\ref{sec:results} tests P1 in Section~\ref{sec:pred}, P2 and P3 in
Section~\ref{sec:hit}, P4 in Section~\ref{sec:imit} and P5 in
Section~\ref{sec:pays}. Equation~\eqref{eq:entry} guides the interpretation of
the returns to entry in Section~\ref{sec:returns}.

\section{Data and methodology}
\label{sec:data}

\subsection{Data}

The data were collected from YouTube. They cover channel and video
characteristics, complete upload histories, view counts, video titles, the tags
creators assign to their videos, full-resolution thumbnail images and speech
transcripts. Transcripts are available for 1,099 of the 1,262 videos in the
feature sample. View counts were recorded between 16 and 20 September 2026.

\subsection{Sampling on birth date}

Most published statistics about creator earnings come from channels a
researcher could find. On YouTube, a channel that fails is almost impossible to
find, so these statistics describe survivors. They show what success looks like
after it has happened, not the odds facing someone about to start.

The central sample is therefore built on the date a channel was born. Broad
daily-life vlog searches, such as ``living alone vlog'', ``day in my life
vlog'' and ``low income daily life'', were run separately
for each of ten monthly windows from November 2025 to August 2026. For every
channel returned, the full upload history was examined, and the channel was
kept only if its first ever upload fell inside the searched month. This rule
depends only on when a channel started, never on how it performed. Among the
channels the searches returned, one with forty lifetime views and one with four
million were equally eligible, provided they started in the same month. The
procedure yields a cold-start cohort of 107
channels.

The cohort captures exactly the outcomes that samples built on visibility miss.
Half of these channels (51.4\%) never reached 1,000 lifetime views, and 11.2\%
stopped after three or fewer uploads.

\subsection{Sample construction}

The content analysis draws on a broader set of channels from two vlog
sub-genres: daily-life vlogs about living on a low income in small rented rooms,
mostly in East and Southeast Asian cities, and vlogs of young adults living
alone in US cities. Candidate channels were identified through keyword searches
for each sub-genre. They were then screened on format and timing only:
long-form videos rather than Shorts, a genre screen applied to titles and
descriptions, and a minimum number of videos where a design requires it.
Performance never enters the screen. Screening on views, subscribers or growth
at any stage would bring back exactly the bias the birth-date frame removes.

\begin{table}[htbp]\centering
\caption{Sample construction. Channels are screened on format and timing only,
never on performance. The rows are not strictly nested. The cold-start cohort is
sampled on birth date and supplies the returns in Table~\ref{tab:ret}. The
event window is the subset with thumbnails and titles collected around each
channel's first breakout.}
\label{tab:sample}
\begin{tabular}{lr}
\toprule
Sample & $N$ \\
\midrule
Channels identified by genre keyword search & 10,911 \\
Channels passing the format and timing screen & 929 \\
Channels with complete upload histories & 420 \\
Videos with publication date and views & 56,419 \\
Videos with extracted content features & 1,262 \\
Cold-start cohort, sampled on birth date & 107 \\
Event-window thumbnails & 2,666 \\
Channels in the event window & 177 \\
\bottomrule
\end{tabular}
\end{table}

Table~\ref{tab:sample} summarises the samples. Keyword searches identify 10,911
distinct channels. Of these, 929 pass the format and timing screen, and 420
channels have complete upload histories. The histories contain 56,419 videos
with a publication date and a view count, and they form the panel behind every
within-channel result. Content features are extracted for a stratified
subsample of 1,262 videos from 319 channels, two top-performing and two
median-performing videos per channel. The event window contains 2,666
thumbnails with their titles and tags from 177 channels, collected in a
symmetric window around each channel's first breakout.

\subsection{A multimodal dataset}

Each video is measured in three modalities: the image that sells it, the speech
that carries it and the text that labels it. Table~\ref{tab:features} lists the
measures. All of them are recorded from the content itself, before and
independently of how the video performed.

\begin{table}[htbp]\centering
\caption{The multimodal feature set. Thumbnail measures come from optical
character recognition and deep learning face detection; speech measures come
from transcripts; title and metadata measures describe each video's title, tags,
description and duration. Of the 37
measures, 36 have enough coverage to enter the predictive models. The lower
panel reports the scale of the data.}
\label{tab:features}
\small
\begin{tabular}{>{\raggedright\arraybackslash}p{2.6cm}>{\raggedright\arraybackslash}p{9.4cm}r}
\toprule
Modality & Measures & Count \\
\midrule
Thumbnail image & Share of the image covered by text, size of the largest
text block, word count, number of text blocks, share of capital letters; number
of faces, size of the largest face, horizontal and vertical position of the
largest face; saturation, brightness, contrast, colourfulness, share of warm
hues, edge density & 15 \\
\addlinespace
Speech & Speaking rate over the whole video and in the first 30 seconds; words
in the first 30 seconds; speed-up of the opening relative to the whole video;
time to the first greeting; questions in the first 60 seconds; frequency of
seven themes (precarity, solitude, domestic life, self-disclosure, direct
address to the viewer, hedging, aspiration); mentions of money; mentions of
numbers & 15 \\
\addlinespace
Title and metadata & Duration, title length, share of capital letters in the
title, number in the title, question in the title, number of tags, length of
the description & 7 \\
\midrule
\multicolumn{3}{l}{\textit{Scale of the data}} \\
\multicolumn{2}{l}{Distinct thumbnails analysed (feature sample and event window)} & 3,728 \\
\multicolumn{2}{l}{Faces detected in feature-sample thumbnails} & 2,260 \\
\multicolumn{2}{l}{Words of on-screen text read from feature-sample thumbnails} & 7,607 \\
\multicolumn{2}{l}{Videos with speech measures} & 1,099 \\
\multicolumn{2}{l}{Hours of speech analysed} & 384.7 \\
\multicolumn{2}{l}{Words transcribed} & 2,940,197 \\
\multicolumn{2}{l}{Videos transcribed with Whisper for validation} & 28 \\
\bottomrule
\end{tabular}
\end{table}

\paragraph{Thumbnails.} A thumbnail is the first thing a viewer sees and the
design element that industry advice emphasises most. Optical character
recognition reads every piece of on-screen text, which gives the share of the
image covered by text, the size of the largest text block, the number of words
and blocks, and the share of capital letters. YuNet, a lightweight deep neural network face detector designed to balance
speed and accuracy \citep{wu2023}, locates the faces in each thumbnail. For each image
the analysis records the number of faces and the size and position of the
largest one. Colour and composition measures describe saturation, brightness,
contrast, colourfulness, the share of warm hues and edge density, and a coarse
histogram of the colour palette supplements them when measuring how similar two
thumbnails look.

\paragraph{Speech.} Speech is where a vlog delivers its content. Transcripts
yield the speaking rate over the whole video and in the opening seconds, how
quickly the creator greets the viewer, how many questions are asked in the first
minute, and how often the creator uses words from seven themes, among them
precarity, solitude, self-disclosure and direct address to the viewer.
The transcripts were validated before use. A subsample of 28 videos was
transcribed independently with the Whisper speech recognition system
\citep{radford2023}, using its large-v3-turbo model, and 12 of the 16 speech measures compared have a
rank correlation of at least 0.90 between the two transcriptions and a median
relative error below 25\%. The validated transcripts allow the speech analysis
to cover 384.7 hours and nearly three million words.
Transcripts were also screened for recognition errors in which software invents
repeated phrases during silence, and such passages were corrected.

\paragraph{Text.} Titles, tags and descriptions record how creators present and
label each video. Tags deserve particular attention. They are not displayed to
viewers, so they record what creators believe a video is about, free of any
attempt to attract clicks.

\paragraph{Interpretable measures.} Every measure is a choice a creator actually
makes, such as a larger face, more text or a warmer palette, rather than a
learned embedding with no direct meaning. When the analysis later measures
imitation, it can therefore say what was imitated. Machine learning serves only
to measure. Every reported effect is estimated by ordinary least squares with
standard errors clustered by channel.

\subsection{Revenue and effort}

Long-form views are converted to advertising revenue using a published global
median RPM (revenue per thousand views) of \$2.30, measured across 300 monetised
channels over 3,595 channel-months \citep{air2026rpm}. RPM is measured after the
platform's 45\% share and includes views that carry no advertising.
Section~\ref{sec:robust} varies the rate from \$1.00 to \$4.00 per thousand
views. Shorts earn far less per view, 3 to 14\% of the long-form rate in almost every
content category
\citep{air2026shorts}, so the baseline counts long-form revenue only and
Section~\ref{sec:robust} adds Shorts revenue explicitly. Effort is measured as
three hours of production per finished minute of video uploaded, Shorts
included. All revenue figures are the advertising revenue implied by observed
views at published rates.

\subsection{Empirical specifications}

\paragraph{Breakouts and event time.} For channel $c$, let $v_{ic}$ be the views
of video $i$, $s_{ic}$ its position among the channel's long-form uploads, and
$m_c$ the channel's median views, computed over long-form videos at least 30
days old. A video is a breakout if $v_{ic} \geq 5\,m_c$, and the channel's first
breakout occurs at position
\begin{equation}
s^{*}_{c} \;=\; \min\left\{\, s_{ic} : v_{ic} \geq 5\,m_c \,\right\}.
\label{eq:breakout}
\end{equation}
Event time $t_{ic} = s_{ic} - s^{*}_{c}$ counts videos relative to the first
breakout, so $t_{ic} = 1$ is the first video after it. Channels that never cross
the threshold form the placebo group.

\paragraph{Predictability.} The outcome is either $y_{ic} = \log_{10} v_{ic}$ or
its deviation from the channel mean, $\tilde{y}_{ic} = y_{ic} - \bar{y}_c$,
which removes each channel's level. Videos are divided into five folds so that
all videos of a channel fall in the same fold. Each model is trained on four
folds and predicts the fifth, and out-of-sample fit is
\begin{equation}
R^{2}_{\text{OOS}} \;=\; 1 \;-\;
\frac{\sum_{i}\bigl(y_i - \hat{y}^{(-k(i))}_{i}\bigr)^2}{\sum_{i}\bigl(y_i - \bar{y}\bigr)^2},
\label{eq:r2}
\end{equation}
where $\hat{y}^{(-k(i))}_{i}$ is the prediction for video $i$ from the model
trained without its fold $k(i)$. Because no model sees a channel's own videos
during training, none can succeed by learning channel levels. Both
gradient-boosted trees and ridge regression are fitted and the better of the
two is reported. This is the demanding choice for a finding of no effect,
because gradient boosting captures nonlinear patterns and interactions
automatically. The same features and folds are used to classify each video's
sub-genre with a gradient-boosted classifier, evaluated by the area under the
ROC curve (AUC), which provides the measurement check in prediction P1.

\paragraph{Similarity to the winning video.} Let $w$ denote channel $c$'s first
breakout video. Visual similarity uses a 66-dimensional description of each
thumbnail, $z_i$, which combines 12 of the 15 thumbnail measures in
Table~\ref{tab:features} with a 54-bin colour histogram (six hue, three
saturation and three brightness levels), each dimension standardised across all
thumbnails. Visual distance is the cosine distance
\begin{equation}
d^{V}_{i} \;=\; 1 \;-\; \frac{z_i \cdot z_w}{\lVert z_i \rVert\,\lVert z_w \rVert}.
\label{eq:dvis}
\end{equation}
Topical distance $d^{T}_{i}$ is defined in the same way from TF-IDF vectors of
video titles, built from single words and word pairs with common English words
removed. Two further topical distances compare sets of words: the content
keywords of the title and the tags the creator assigns. For each,
\begin{equation}
d^{K}_{i} \;=\; 1 \;-\; \frac{\lvert K_i \cap K_w \rvert}{\lvert K_i \cup K_w \rvert},
\label{eq:jaccard}
\end{equation}
where $K_i$ is the set of keywords or tags of video $i$. Similarity is one minus
distance.

\paragraph{The effect of a hit.} For each channel with at least ten videos and at
least three videos on each side of its first breakout, an interrupted time series
is estimated in upload order:
\begin{equation}
\log_{10} v_{ic} \;=\; \alpha_c \;+\; \beta_{\text{seq}}\, s_{ic}
\;+\; \beta_{\text{post}}\, \mathbf{1}[s_{ic} > s^{*}_{c}] \;+\; \varepsilon_{ic},
\label{eq:its}
\end{equation}
where $\alpha_c$ is a channel fixed effect and standard errors are clustered by
channel. Equation~\eqref{eq:its} is the empirical counterpart of
Equation~\eqref{eq:model}: $\beta_{\text{post}}$ estimates the lasting gain from
a hit, $\delta$, and $\beta_{\text{seq}}$ estimates the return to experience,
$\lambda$. The implied percentage change in views after a hit is
$100\,(10^{\beta_{\text{post}}} - 1)$.

Three features of the design matter. The breakout video itself is excluded from
both periods, so the estimate reflects later videos rather than the hit. The
sequence term absorbs normal channel growth, so ordinary ageing is not mistaken
for the effect of a hit. Videos published within 30 days of data collection are
dropped, because they are still gaining views. The placebo assigns each channel
in the placebo group a false breakout at the median position of its uploads and
estimates the same equation.

\paragraph{Imitation.} Let $d_{ic}$ be the distance of video $i$ from its
channel's winning video, in either dimension, and let
$\text{Post}_{ic} = \mathbf{1}[t_{ic} > 0]$. The baseline estimates the change in
distance after the breakout within each channel,
\begin{equation}
d_{ic} \;=\; \alpha_c \;+\; \beta\, \text{Post}_{ic} \;+\; \varepsilon_{ic},
\label{eq:pooled}
\end{equation}
so a negative $\beta$ means creators move toward the winner. Robustness
specifications add the distance in time from the breakout, $\lvert t_{ic}
\rvert$, or estimate a trend before the breakout together with a change in slope
after it,
\begin{equation}
d_{ic} \;=\; \alpha_c \;+\; \eta\, t_{ic} \;+\; \beta\, \text{Post}_{ic}
\;+\; \kappa\,\bigl(\text{Post}_{ic} \times t_{ic}\bigr) \;+\; \varepsilon_{ic},
\label{eq:slope}
\end{equation}
in which $\eta$ is the trend before the breakout, $\beta$ the jump at the
breakout and $\kappa$ the change in slope afterwards. The event study estimates a
separate coefficient for each position,
\begin{equation}
d_{ic} \;=\; \alpha_c \;+\; \sum_{\tau \neq -1} \gamma_\tau\,
\mathbf{1}[t_{ic} = \tau] \;+\; \varepsilon_{ic},
\qquad \tau \in [-8, +8],
\label{eq:event}
\end{equation}
with $t=-1$ as the reference period. The placebo re-estimates
Equation~\eqref{eq:pooled} with distance measured to a randomly chosen video from
the same channel other than the winner. Prediction P4 implies $\beta < 0$ in both
dimensions and a drop in $\gamma_\tau$ at the breakout.

\paragraph{Returns to imitation.} For videos published after the breakout
($t_{ic} > 0$), in channels with at least three such videos,
\begin{equation}
\log_{10} v_{ic} \;=\; \alpha_c \;+\; \pi_T\, \tilde{S}^{T}_{ic}
\;+\; \pi_V\, \tilde{S}^{V}_{ic} \;+\; u_{ic},
\label{eq:pays}
\end{equation}
where $\tilde{S}^{T}_{ic}$ and $\tilde{S}^{V}_{ic}$ are topical and visual
similarity to the winner, standardised to mean zero and unit variance.
Prediction P5 holds if $\pi_T > 0$ while $\pi_V = 0$, the empirical counterpart
of $\theta_T > 0$ and $\theta_V = 0$ in Equation~\eqref{eq:model}.

\paragraph{Returns to entry.} For each channel in the cold-start cohort, implied
revenue, production hours and the hourly return are
\begin{equation}
R_c \;=\; \frac{V^{L}_{c}}{1000}\,\text{RPM}, \qquad
H_c \;=\; 3\,M_c, \qquad
w_c \;=\; \frac{R_c}{H_c},
\label{eq:returns}
\end{equation}
where $V^{L}_{c}$ is the channel's lifetime long-form views and $M_c$ the total
minutes of video it has uploaded. Inequality is summarised by the Gini
coefficient,
\begin{equation}
G \;=\; \frac{\sum_{j=1}^{n}\,(2j - n - 1)\,R_{(j)}}{n\,\sum_{j=1}^{n} R_{(j)}},
\label{eq:gini}
\end{equation}
where $R_{(1)} \leq \cdots \leq R_{(n)}$ are channel revenues in ascending order.

\subsection{Identification strategy}

The predictability result is a well-powered finding of no effect and requires no
causal assumption. The imitation result comes from a within-channel event study.
It is identified by the sharp change at the breakout rather than by levels, and
its key test is a placebo that measures similarity to a randomly chosen other
video from the same channel.

The effect of a hit is estimated by an interrupted time series. Because
breakouts are not randomly assigned, the design is supported by three checks: a
placebo that assigns a false breakout to channels that never had one, 17
alternative specifications in Section~\ref{sec:robust}, and an analysis of the
trend before the breakout in Section~\ref{sec:threats}. The returns estimates
describe revenue implied by observed views.

\section{Results}
\label{sec:results}

Of the total variation in $\log_{10}$ views across the 1,262 feature-extracted
videos, \textbf{80.2\%} lies between channels and only 19.8\% within them. A
video's audience depends mostly on the channel that published it rather than on
the video itself. The key
questions are therefore what sets a channel's level, and whether anything a
creator does can change it.

\subsection{Content does not predict success}
\label{sec:pred}

\begin{table}[htbp]\centering
\caption{Out-of-sample predictive performance of content features. Folds are
split by channel, so no model sees a channel's own videos during training. Each
cell reports the better of gradient-boosted trees and ridge regression. Negative
out-of-sample $R^2$ values are reported as zero. The last row applies the same
features and folds to classifying each video's sub-genre.}
\label{tab:pred}
\begin{tabular}{lcc}
\toprule
 & \multicolumn{2}{c}{Out-of-sample $R^2$} \\
\cmidrule(lr){2-3}
Predictors & Raw log views & Within channel \\
\midrule
Thumbnail features & 0.006 & 0.000 \\
Speech features & 0.000 & 0.004 \\
Title and metadata & 0.015 & 0.001 \\
All content features & 0.039 & 0.007 \\
\quad + video age control & 0.145 & 0.011 \\
\midrule
\textit{Validity check:} genre & \multicolumn{2}{c}{AUC $= 0.782$} \\
\bottomrule
\end{tabular}
\end{table}

Table~\ref{tab:pred} tests prediction P1. Content features explain almost none
of the variation in views within a channel. The best specification reaches an
$R^2$ of 0.011, and only with a control for video age. Thumbnail features,
which industry advice treats as the main driver of clicks, reach 0.000.

The raw-views column is higher, but this reflects differences between channels
rather than content. A model given only a video's age reaches an $R^2$ of 0.064
on raw views, which is most of what the full set of content features achieves.
Once channel identity is removed, the predictive power disappears with it.

\begin{figure}[htbp]\centering
\includegraphics[width=0.85\textwidth]{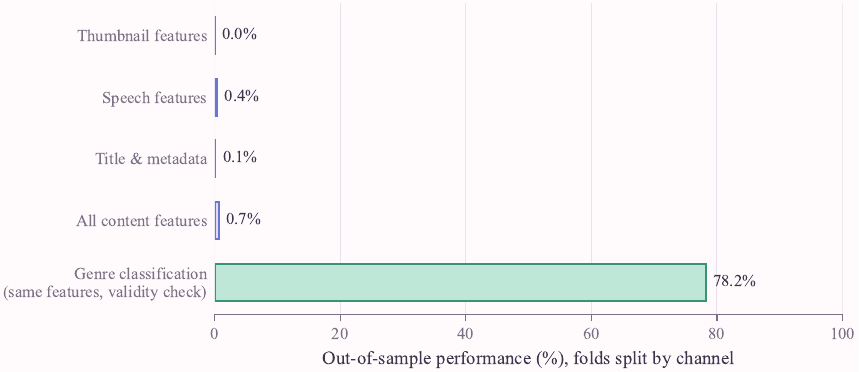}
\caption{Content features predict sub-genre but not success. The first four bars
show the within-channel out-of-sample $R^2$ for predicting views from each group
of features, as a percentage. The last bar shows the AUC for classifying a
video's sub-genre from all features combined. The same measurements that
classify sub-genre with an AUC of 0.782 explain almost none of the variation in
views.}
\label{fig:pred}
\end{figure}

\paragraph{The measures carry real information.} A finding of no effect is meaningful only if
the measurement is sound. Figure~\ref{fig:pred} shows that it is. Using the same
features and the same channel-split folds, a gradient-boosted classifier
distinguishes the two vlog sub-genres with an AUC of 0.782 across 1,262 videos
from 319 channels. The features contain real, recoverable information about what
a video is. They contain none about what it earns.

Two further tests point the same way. Distance from the typical video in the
genre has no relationship with views within a channel, either as a straight
line ($-0.026$, $t=-0.72$) or as a curve ($+0.018$, $t=+0.81$). Standing out
does not pay, fitting in does not pay, and there is no sweet spot in between.
Channels also do not move toward the visual style of the genre's most successful
videos as they age ($+0.0019$ per video, $t=+0.26$). Whatever separates winners
from losers, it is not in the content.

\subsection{Nothing accumulates except a hit}
\label{sec:hit}

\begin{table}[htbp]\centering
\caption{Effect of a channel's first breakout on its later videos, from
Equation~\eqref{eq:its}. A breakout is a video with at least five times its
channel's median views. The breakout video is excluded from both periods. The
outcome is $\log_{10}$ views, with channel fixed effects and standard errors
clustered by channel in parentheses. The placebo column assigns a false breakout
at the midpoint of channels that never had one. $^{***}p<0.01$.}
\label{tab:spill}
\begin{tabular}{lcc}
\toprule
 & Treated & Placebo \\
 & (first breakout) & (never-breakout) \\
\midrule
Post-breakout & +0.4219$^{***}$ & +0.0339 \\
 & (0.0553) & (0.0313) \\
 & [$t=7.63$] & [$t=1.08$] \\
Upload-sequence trend & $-0.00038$ & $-0.00058$ \\
\midrule
Implied change in views & +164.2\% & +8.1\% \\
Videos & 21,605 & 1,196 \\
Channels & 159 & 22 \\
\bottomrule
\end{tabular}
\end{table}

Table~\ref{tab:spill} tests predictions P2 and P3 and contains the largest
effect in the paper. A channel's first breakout raises the views of its later
videos by $+0.4219$ in $\log_{10}$ terms, a factor of 2.64, or \textbf{164\%}.
This is the estimate of $\delta$ in Equation~\eqref{eq:audience}. The placebo
runs the same model on channels that never broke out, using a false breakout at
the midpoint of their history. It finds $+8.1\%$ with a $t$-statistic of 1.08,
which is not significant. The design does not create jumps where none exist.

Experience has no such effect. The estimate of $\lambda$, the
upload-sequence coefficient, is $-0.00038$ per video, so forty more videos
change views by $-3.4\%$. Producing more, without a hit, does not raise
performance. Learning by doing \citep{arrow1962} is absent even though the
setting is built for it: the task repeats, feedback is
immediate and numerical, and creators are highly motivated. The only thing that
moves a channel's position is a hit.

\begin{figure}[htbp]\centering
\includegraphics[width=0.82\textwidth]{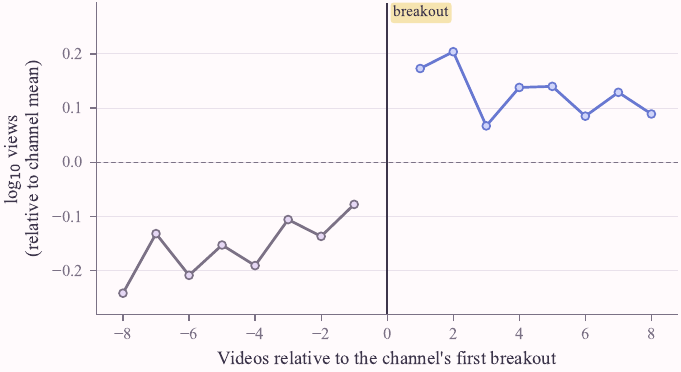}
\caption{Views before and after a channel's first breakout, relative to the
channel's own average. The horizontal axis counts videos, not days. The vertical
line marks the breakout, which is excluded. Every point before the breakout lies
below the channel average and every point after it lies above, with the jump
occurring exactly at the breakout. Within this window of eight videos on each
side, the shift is $+92\%$. Table~\ref{tab:spill} reports $+164\%$ because it
uses every video after the breakout.}
\label{fig:spill}
\end{figure}

Figure~\ref{fig:spill} shows the shift and its timing. From the video just
before the hit to the video just after, views rise by 78\%. The level then stays
high for all eight later videos observed, easing only slightly ($-0.011$ per
video). Views also rise gently in the videos leading up to the breakout.
Section~\ref{sec:threats} shows that this earlier trend accounts for only a
small part of the jump.

\begin{figure}[htbp]\centering
\includegraphics[width=0.88\textwidth]{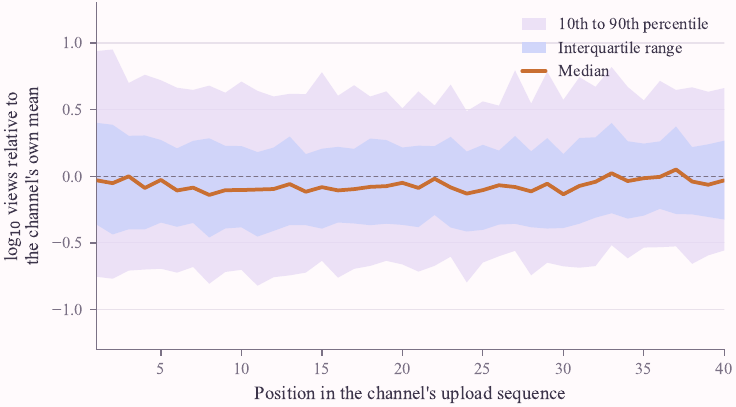}
\caption{Within-channel performance by position in the upload sequence. Each
video's views are measured relative to its own channel's average, so
differences in channel size, audience and niche drop out. The bands show the
interquartile range and the 10th to 90th percentile range across channels at
each position. The median is flat, so a channel's fortieth video does no better
than its first. The spread does not narrow, so later videos are no more
predictable either. Experience improves neither the level nor the consistency
of results.}
\label{fig:experience}
\end{figure}

Figure~\ref{fig:experience} shows the absence of learning directly. The
comparison has to be made within channels. Across channels, those with more
videos do have more views per video, but that pattern reflects survival rather
than learning. Creators who fail stop posting, so channels with long histories
are the ones that did well. Within each channel, performance does not improve
with experience.

\begin{figure}[htbp]\centering
\includegraphics[width=0.62\textwidth]{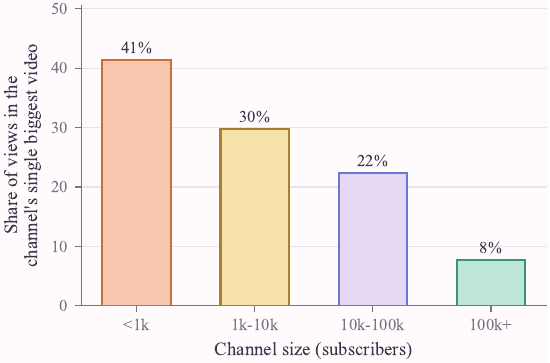}
\caption{Share of a channel's lifetime views that comes from its single
most-viewed video, by channel size, across complete upload histories. The share
falls steadily from 41.4\% to 7.7\%. Channels that have not had a hit are
defined by their one best video. Channels that have had one are not.}
\label{fig:conc}
\end{figure}

\paragraph{The same pattern across channels.} If a hit sets a channel's level,
small channels should depend on a single video and large ones should not.
Figure~\ref{fig:conc} confirms this. The smallest channels draw
\textbf{41.4\%} of all their lifetime views from one video. The share falls
steadily to \textbf{7.7\%} for the largest channels. A small channel is one
video. A large channel has a floor.

\subsection{Creators copy their own hit}
\label{sec:imit}

If a hit is the only thing that moves a channel's position, and creators notice
this, they should respond when one arrives. Prediction P4 says they should move
toward the winner in every dimension at once, and they do.

\begin{figure}[htbp]\centering
\includegraphics[width=0.88\textwidth]{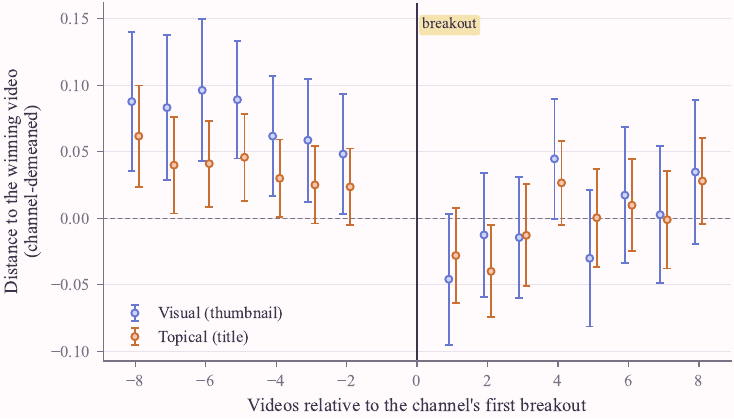}
\caption{Distance to the winning video by position relative to a channel's first
breakout, from Equation~\eqref{eq:event}. The visual measure uses thumbnail
features and the topical measure uses the TF-IDF vectors of titles.
Coefficients are relative to $t=-1$, and bars show 95\% confidence intervals
clustered by channel. Lower values mean more similar to the winner. Both series
lie above zero before the breakout and fall below zero immediately after it.
They move together even though they share no data: the visual measure never
sees a title, and the topical measure never sees an image.}
\label{fig:event}
\end{figure}

Figure~\ref{fig:event} shows the central behavioural result. After a breakout,
creators move toward the winning video in both dimensions at once. The agreement
between the two measures is the strongest part of the result. A visual shift
alone could reflect a rebranding, and a topical shift alone could reflect a
narrowing of subject matter for other reasons. Two measures built from different
data, moving together at the same moment, point clearly to imitation.

\begin{table}[htbp]\centering
\caption{Imitation of the winning video across specifications. The outcome is
distance to the breakout video, so a negative coefficient means movement toward
the winner. All specifications include channel fixed effects, with standard
errors clustered by channel. The last row is a placebo that measures distance
to a randomly chosen video from the same channel. Blank cells are
specifications not estimated for that measure. $^{**}p<0.05$,
$^{***}p<0.01$.}
\label{tab:imit}
\begin{tabular}{lcc}
\toprule
 & Visual & Topical \\
Specification & (thumbnail) & (title TF-IDF) \\
\midrule
Baseline & $-0.0612^{***}$ & $-0.0351^{***}$ \\
Excluding $t=-1$ & $-0.0726^{***}$ & \\
Donut (excl.\ $t=\pm1$) & $-0.0649^{***}$ & \\
$+$ temporal proximity & $-0.0651^{***}$ & $-0.0382^{***}$ \\
$+$ trend and slope change & $-0.0486^{**}$ & $-0.0413^{**}$ \\
\textit{Placebo:} random own video & & $+0.0040$ \\
\midrule
Channels & 177 & 177 \\
Observations & 2,489 & 2,312 \\
\bottomrule
\end{tabular}
\end{table}

Table~\ref{tab:imit} shows that the result does not depend on any single
specification. Dropping the reference period, removing the videos right next to
the breakout, controlling for how close videos are in time, and allowing for
both a trend before the breakout and a change in slope after it all leave the
result intact. The visual coefficient ranges from $-0.049$ to $-0.073$ and the
topical coefficient from $-0.035$ to $-0.041$.

\paragraph{The placebo.} Creators often narrow their subject matter over time as
they find an audience or specialise. That would bring every video closer to
every other video in the channel. Measured against a randomly chosen other
video from the channel, the coefficient is $+0.0040$ ($t=0.43$), effectively zero. Creators move
toward the winner specifically, not toward their own past work in general.

\begin{table}[htbp]\centering
\caption{Topical imitation under three independent measures of subject matter,
with placebos. Titles are written to attract clicks. Tags are not displayed to
viewers and therefore record what the creator believes the video is about. All
three measures show movement toward the winner and neither placebo does.
$^{**}p<0.05$, $^{***}p<0.01$.}
\label{tab:topical}
\begin{tabular}{lcccc}
\toprule
Measure of topical distance & Coefficient & $t$ & $N$ & Channels \\
\midrule
Title TF-IDF cosine & $-0.0351^{***}$ & $-4.10$ & 2,312 & 177 \\
Content-keyword Jaccard & $-0.0228^{***}$ & $-3.94$ & 2,290 & 176 \\
Creator-tag Jaccard & $-0.0189^{**}$ & $-2.55$ & 1,592 & 131 \\
\midrule
\textit{Placebo:} TF-IDF to random own video & $+0.0040$ & $+0.43$ & 2,312 & 177 \\
\textit{Placebo:} keywords to random own video & $+0.0038$ & $+0.68$ & 2,271 & 174 \\
\bottomrule
\end{tabular}
\end{table}

Table~\ref{tab:topical} repeats the topical test with three different
definitions of subject matter. All three show movement toward the winner, and
both placebos are zero. The tag result is the most convincing of the three.
Viewers do not see tags, so a
shift in tags cannot be explained as copying a title style that happened to
work. It shows that the creator's own classification of their work moves toward
the hit.

\paragraph{A jump, then a fade.} When a trend before the breakout and a change
in slope after it are estimated together, both measures show the pattern that
Section~\ref{sec:theory} predicts: gradual convergence beforehand, a sharp jump
at the breakout, and a reversal afterwards (visual $+0.0200$, $t=+4.20$; topical
$+0.0154$, $t=+4.56$). Imitation is a response to a shock, not a permanent
change of style.

\subsection{Only topical imitation pays}
\label{sec:pays}

\begin{table}[htbp]\centering
\caption{Returns to resembling the winner. The sample includes only videos
published after the breakout, from channels with at least three such videos, so
each creator had a winner available to copy. The outcome is $\log_{10}$ views.
Both similarity measures are standardised, so coefficients are per standard
deviation. All columns include channel fixed effects, with standard errors
clustered by channel in parentheses. $^{**}p<0.05$.}
\label{tab:pays}
\begin{tabular}{lccc}
\toprule
 & (1) & (2) & (3) \\
\midrule
Topical similarity to winner & $+0.0635^{**}$ & & $+0.0617^{**}$ \\
 & (0.0251) & & (0.0252) \\
Visual similarity to winner & & $+0.0386$ & $+0.0350$ \\
 & & (0.0243) & (0.0247) \\
\midrule
Channel fixed effects & Yes & Yes & Yes \\
Observations & 1,336 & 1,336 & 1,336 \\
Channels & 177 & 177 & 177 \\
\bottomrule
\end{tabular}
\end{table}

Table~\ref{tab:pays} tests prediction P5 by asking whether the copying in
Section~\ref{sec:imit} paid off. A one standard deviation increase in topical
similarity to the winner raises views by about 16\%, and the effect is
statistically significant. A one standard deviation increase in visual
similarity has an estimated effect of 0.0386 with a standard error of 0.0243,
which is not statistically different from zero. Entering both measures together
in column (3) leaves both results unchanged.

Creators copy both the topic and the look, but only the topic pays. In the terms of
Section~\ref{sec:theory}, topic carries demand ($\theta_T > 0$) and design does
not ($\theta_V = 0$), yet creators imitate both, exactly as a single confounded
signal implies.

This also clarifies the source of the 164\% effect in Section~\ref{sec:hit}.
The effect comes from the audience a hit brings rather than from imitation, and
it applies to later videos whether or not they resemble the winner. Matching the
winner's topic adds a further gain on top of it, while matching its look adds
nothing.

\subsection{Returns to entry}
\label{sec:returns}

\begin{table}[htbp]\centering
\caption{Returns to entry for the cold-start cohort. Channels are sampled on
birth date rather than visibility, so the distribution includes channels that
failed. Revenue converts lifetime long-form views at a published global median RPM
of \$2.30 per thousand views. Hours assume three hours of production per
finished minute of video uploaded, Shorts included. Section~\ref{sec:robust}
varies the rate and adds Shorts revenue.}
\label{tab:ret}
\begin{tabular}{lr}
\toprule
Cold-start cohort ($N = 107$) & \\
\midrule
Median total revenue & \$1.61 \\
75th percentile & \$61.74 \\
90th percentile & \$765.94 \\
Share earning $<$ \$100 & 79.4\% \\
Share earning $<$ \$1{,}000 & 90.7\% \\
Gini of earnings & 0.906 \\
Top decile share of cohort revenue & 87.9\% \\
\midrule
Median hours invested & 185 \\
Median implied hourly return & \$0.01 \\
Share below US federal minimum wage & 97.2\% \\
\bottomrule
\end{tabular}
\end{table}

Table~\ref{tab:ret} reports the advertising revenue implied by the views of
the 107 channels followed from their first upload. The median channel's
long-form views were worth \textbf{\$1.61}, against an estimated \textbf{185
hours} of work. For 97.2\% of the cohort, the implied
hourly return was below the US federal minimum wage. Earnings are extremely
unequal. The Gini coefficient is 0.906, and the top tenth of channels captured
87.9\% of all the revenue the cohort generated.

\begin{figure}[htbp]\centering
\includegraphics[width=0.55\textwidth]{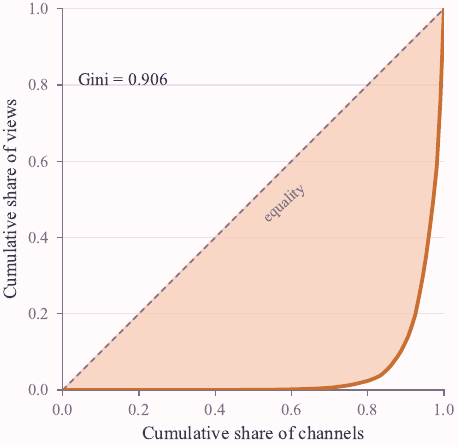}
\caption{Lorenz curve of lifetime views across the cold-start cohort, against
the 45-degree line of perfect equality. Because the cohort is selected on birth
date rather than visibility, the curve includes the channels that failed, not
only those that survived. It therefore lies much further from equality than published
creator-earnings figures, which are based on channels a researcher could
find.}
\label{fig:lorenz}
\end{figure}

Figure~\ref{fig:lorenz} plots the full distribution. Its shape is familiar from
superstar markets. What is new is the sample. Because channels enter the frame
based on when they started rather than whether they succeeded, this curve shows
the prospects facing someone deciding whether to start. That is the
distribution an entrant needs, and one that earlier studies could not observe.
Equation~\eqref{eq:entry} explains why such low median returns are compatible
with continued entry: the value of entry lies in the lasting prize rather than
in the typical outcome.

\section{Robustness}
\label{sec:robust}

This section tests whether the conclusions depend on choices made in the
analysis. Several results rely on a threshold, a time window, a sample
restriction or a functional form. Each of these choices is varied for four
results: the effect of a hit, imitation, the predictability finding and the
returns to entry. All four hold under every variation.

\subsection{The effect of a hit}

The estimate in Section~\ref{sec:hit} depends on four choices: the multiple of
the channel median that defines a breakout, the time window that excludes videos
still gaining views, the minimum number of videos per channel, and the
straight-line form of the upload-sequence trend. The breakout threshold matters
most because it defines the event itself. Table~\ref{tab:robust} varies all
four, together with trimming extreme values of the outcome, in 17 alternative
specifications alongside the baseline.

\begin{table}[htbp]\centering
\caption{Robustness of the effect of a hit. Each row re-estimates
Equation~\eqref{eq:its} under the stated choice, holding the others at their
baseline values. The baseline (five times the median, 30 days, at least 10
videos, at least 3 videos on each side, linear trend) reproduces
Table~\ref{tab:spill}. Standard errors are clustered by channel. All 18
coefficients are positive and significant at the 1\% level. $^{***}p<0.01$.}
\label{tab:robust}
\begin{tabular}{lcccr}
\toprule
Specification & $\beta_{\text{post}}$ & $t$ & Implied & Videos \\
\midrule
\multicolumn{5}{l}{\textit{Panel A. Breakout threshold}} \\
\quad $2\times$ channel median & $+0.5655^{***}$ & $+10.10$ & $+267.7\%$ & 15,929 \\
\quad $3\times$ channel median & $+0.5131^{***}$ & $+8.51$ & $+225.9\%$ & 19,186 \\
\quad $4\times$ channel median & $+0.4669^{***}$ & $+8.40$ & $+193.0\%$ & 20,815 \\
\quad $\mathbf{5\times}$ \textbf{channel median (baseline)} & $+0.4219^{***}$ & $+7.63$ & $+164.2\%$ & 21,605 \\
\quad $7\times$ channel median & $+0.4008^{***}$ & $+8.21$ & $+151.6\%$ & 22,485 \\
\quad $10\times$ channel median & $+0.4041^{***}$ & $+7.84$ & $+153.6\%$ & 21,806 \\
\addlinespace
\multicolumn{5}{l}{\textit{Panel B. Maturity window}} \\
\quad No maturity filter & $+0.4192^{***}$ & $+7.92$ & $+162.5\%$ & 22,453 \\
\quad Drop videos $<$ 60 days old & $+0.4237^{***}$ & $+7.40$ & $+165.3\%$ & 20,942 \\
\quad Drop videos $<$ 90 days old & $+0.4302^{***}$ & $+7.30$ & $+169.3\%$ & 20,393 \\
\quad Drop videos $<$ 180 days old & $+0.4316^{***}$ & $+6.89$ & $+170.1\%$ & 19,055 \\
\addlinespace
\multicolumn{5}{l}{\textit{Panel C. Sample restrictions}} \\
\quad Channels with $\geq 15$ videos & $+0.4267^{***}$ & $+7.63$ & $+167.1\%$ & 21,505 \\
\quad Channels with $\geq 20$ videos & $+0.4303^{***}$ & $+7.59$ & $+169.4\%$ & 21,377 \\
\quad Channels with $\geq 30$ videos & $+0.4386^{***}$ & $+7.46$ & $+174.6\%$ & 21,013 \\
\quad $\geq 5$ videos each side of event & $+0.4503^{***}$ & $+7.81$ & $+182.1\%$ & 17,538 \\
\quad $\geq 8$ videos each side of event & $+0.4650^{***}$ & $+7.38$ & $+191.7\%$ & 13,700 \\
\addlinespace
\multicolumn{5}{l}{\textit{Panel D. Outcome and functional form}} \\
\quad Winsorised at 1st/99th pct & $+0.4151^{***}$ & $+7.55$ & $+160.1\%$ & 21,605 \\
\quad Winsorised at 5th/95th pct & $+0.3836^{***}$ & $+7.09$ & $+141.9\%$ & 21,605 \\
\quad Quadratic sequence trend & $+0.4092^{***}$ & $+6.98$ & $+156.6\%$ & 21,605 \\
\bottomrule
\end{tabular}
\end{table}

The estimate is positive and significant at the 1\% level in all 18
specifications. The smallest $t$-statistic is 6.89, more than three times the
usual threshold. The implied effect ranges from $+142\%$ to $+268\%$. Panels B
to D change the estimate very little. Removing the maturity filter or tightening
it to 180 days, keeping only channels with at least 15, 20 or 30 videos,
requiring at least five or eight videos on each side of the breakout, trimming
extreme values and allowing the trend to curve all keep the effect between
$+142\%$ and $+192\%$.

Panel A shows how the estimate responds to the breakout threshold. It falls
steadily as the threshold rises, from $+268\%$ at twice the median to about
$+152\%$ at seven to ten times the median, where it settles. A low threshold
treats ordinary variation as a breakout and detects it early in a channel's
history, so part of the channel's normal growth is counted as the effect of a
hit. The higher thresholds give the cleanest comparison, and even there a first
breakout raises later views by more than 150\%. The effect of a hit is large
under every definition.

Every specification in Table~\ref{tab:robust} rests on the same assumption about
the timing of breakouts. Section~\ref{sec:threats} examines that assumption
directly.

\subsection{Imitation}

The imitation results carry their own robustness checks in
Section~\ref{sec:imit}. Across the five specifications in
Table~\ref{tab:imit}, visual movement toward the winner ranges from $-0.049$ to
$-0.073$ and topical movement from $-0.035$ to $-0.041$, all significant at the
5\% level or better. The most demanding specification, which allows a trend
before the breakout and a change in slope after it, still finds the jump at the
breakout, with $t=-2.48$ for the visual measure and $t=-2.49$ for the topical
measure.

Table~\ref{tab:topical} shows that the topical result does not depend on how
subject matter is measured. Titles, content keywords and creator tags are three
measures with different weaknesses. All three show movement toward the winner,
and both placebos are zero.

\subsection{Predictability}

A finding of no effect is robust when no reasonable modelling choice would have
found one. Table~\ref{tab:pred} reports each result as the better of a linear
model and a flexible one, for two definitions of the outcome, with folds split by
channel so that no model can succeed by memorising a channel's typical level.
The within-channel $R^2$ never exceeds 0.011. The possibility that the effect of
a design choice is curved, with a sweet spot that a straight line would miss, is
tested directly. Distance from the genre's typical video has no relationship
with views, either linear ($t=-0.72$) or quadratic ($t=+0.81$). The sub-genre
benchmark confirms that the same features and folds detect real patterns when
they exist (AUC 0.782).

\subsection{Revenue conversion}

Table~\ref{tab:rpm} varies the long-form rate from \$1.00 to \$4.00 per
thousand views. The Gini coefficient does not depend on the rate, so it stays at
0.906 throughout, and the shares of channels below fixed dollar thresholds move
only modestly. Even at the highest rate, the median channel's long-form views
are worth \$2.81, and 96.3\% of the cohort earns less per hour than the minimum
wage.

Shorts deserve separate attention, because they account for 81.3\% of the
cohort's views. Adding Shorts revenue at \$0.15 per thousand Shorts views, about
6.5\% of the long-form median and inside the measured range of 3 to 14\%
\citep{air2026shorts}, raises the median to \$9.08 and lowers the Gini
coefficient to 0.890. The conclusion is unchanged: 96.3\% of the cohort still
earns less per hour than the minimum wage. The finding that entry pays almost
nothing does not depend on the revenue assumptions.

\begin{table}[htbp]\centering
\caption{Sensitivity of returns to the revenue assumptions. The long-form rate
varies around the measured median of \$2.30 used in Table~\ref{tab:ret}; the
last row adds Shorts revenue at \$0.15 per thousand Shorts views. The Gini coefficient does not depend on the long-form rate; it
changes when Shorts are added because channels differ in how much of their
viewing comes from Shorts. The shares below \$100 and below the minimum wage change only because those dollar
thresholds are fixed.}
\label{tab:rpm}
\begin{tabular}{lrrrrrr}
\toprule
Assumption & Median & 75th pct & 90th pct & $<$\$100 & $<$ min wage & Gini \\
\midrule
\$1.00 (low) & \$0.70 & \$26.84 & \$333.02 & 83.2\% & 100.0\% & 0.906 \\
\textbf{\$2.30 (baseline)} & \textbf{\$1.61} & \$61.74 & \$765.94 & 79.4\% & 97.2\% & 0.906 \\
\$4.00 (high) & \$2.81 & \$107.37 & \$1{,}332.07 & 74.8\% & 96.3\% & 0.906 \\
\midrule
\$2.30 plus Shorts at \$0.15 & \$9.08 & \$99.01 & \$1{,}079.50 & 74.8\% & 96.3\% & 0.890 \\
\bottomrule
\end{tabular}
\end{table}

\section{Discussion}
\label{sec:disc}

\subsection{Validity of the design}
\label{sec:threats}

\paragraph{The trend before a breakout.} Views rise gently in the eight videos
before a breakout, by 0.0166 in $\log_{10}$ terms per video, with a correlation
of $+0.74$ between views and position over those eight videos.
Equation~\eqref{eq:its} already controls for a linear trend over each channel's
full history, but the local trend just before a breakout is steeper. The
question is whether this trend could explain the jump. Extending the
pre-breakout trend through the event, the immediate jump is still $+65\%$,
compared with $+78\%$ without the adjustment. Two further facts show that the
jump is not a continuation of earlier growth: channels in the placebo group show
no comparable rise, and the change happens at a single upload, which gradual
improvement would not produce.

\paragraph{The trend before imitation.} Creators were already moving slightly
toward the eventual winner's style before it succeeded. Splitting the
pre-breakout period at its midpoint gives $-0.049$ ($t=-3.48$). Part of this is
mechanical, since style changes gradually and videos made close together
resemble each other. The jump at the breakout remains after controlling for
this, because the change is asymmetric: videos after the breakout are closer to
the winner than videos the same distance before it. With a pre-breakout trend
and a post-breakout change in slope estimated together, the jump remains
significant ($t=-2.48$).

\paragraph{The timing of breakouts.} Breakouts are not randomly assigned, so the
design identifies a sharp change at the event rather than an experimentally
assigned effect. Three features of the evidence support interpreting it as the
effect of the hit. The change is concentrated at a single upload, it is absent
in placebo channels, and it holds across all 17 alternative specifications.
Section~\ref{sec:conc} describes a design that isolates a single cause.

\paragraph{Production choices and quality.} The features deliberately measure
production choices rather than an abstract notion of quality, because these are
the choices creators control and that industry advice promotes: face size,
on-screen text, colour, speaking pace and topic. None of them predicts returns
within a channel, even though together they identify sub-genre with an AUC of
0.782. For talent to explain success, it would have to lie in qualities that
leave no trace in any of these characteristics.

\paragraph{Audience and algorithm.} A hit's lasting effect could work through
returning viewers, through the recommendation system, or through higher visible
view counts that attract more clicks. On the platform these forces operate
together, and the estimate captures their combined effect, which is the quantity
$\delta$ in Equation~\eqref{eq:audience}. Separating them requires
impression-level data held by the platform, which makes this a natural direction
for research in partnership with platforms.

\subsection{Why people enter: a lasting prize}

Table~\ref{tab:ret} poses a classic puzzle. New creators invest
about 185 hours for a median return of \$1.61. Entry into occupations with poor
average returns is well documented \citep{hamilton2000,moskowitz2002}, and the
usual explanations rely on non-financial rewards or overconfidence
\citep{camerer1999,astebro2014}.

The framework in Section~\ref{sec:theory} points to an explanation that requires
neither. What a new creator buys is a chance at a lasting prize rather than one
video's revenue, since a hit raises the views of everything published afterwards
by 164\%. Equation~\eqref{eq:entry} shows
that the value of this prize grows with the number of videos a creator expects to
make, so it can far exceed the typical outcome. Entering a market with a median
return of \$1.61 can therefore be rational, as long as the creator plans to keep
publishing. The argument complements models in which early success reveals
ability \citep{macdonald1988}: here the prize persists because the audience
persists, and the content evidence gives no sign that a hit reveals any
measurable skill.

This explanation can be tested. If entry reflects overconfidence, creators who
learn the true odds should leave. If it reflects a lasting prize, they should
stay, because the odds for any single video were never the relevant quantity.

\subsection{Why creators copy what does not pay}

Read together, the imitation results and their returns show that creators copy
a hit in two ways, and only one of them pays. Copying the
look resembles what \citet{levitt1988} describe as superstitious learning:
crediting an outcome to a feature that did not cause it.

Creators are not simply making a mistake, given what they can see. A hit is a
single event in which the topic, the thumbnail, the timing, the algorithm and
chance all combine. The platform reports far more than views. Creators see
impressions, click-through rates and watch time \citep{youtube2026reach} and how
well each part of a video holds its audience \citep{youtube2026retention}, and
those who enable advanced features can test up to three titles and thumbnails
against each other \citep{youtube2026abtest}. None of this separates the causes
of a hit. The statistics describe how a hit performed without explaining why;
a high click-through rate can reflect an appealing thumbnail or a topic in demand. A
test compares packaging on a single video. It cannot vary the topic, it excludes
Shorts, and like any experiment it needs traffic, which small channels lack. Testing a topic still means publishing videos that may perform
worse. Equation~\eqref{eq:posterior} makes the point precise: after a single
success, the update for each feature of the hit depends on how distinctive the
feature was rather than on whether it mattered.

When the only evidence is one mixed signal from one event, copying every visible
feature is the sensible response. It is the same inference problem that this
paper needed 2,489 video comparisons and an event study to answer. The effort
spent copying the look is real and earns nothing, yet it stems from the
information creators receive rather than from poor judgement.

The topical result is consistent with two mechanisms. Imitation itself may work,
or audiences may want more of a subject that has already succeeded, as readers
of news do \citep{ho2015}. Both make
copying the topic a good strategy, and neither changes the finding that copying
the look does not pay.

\subsection{Rosen, Adler, and what the evidence shows}

The superstar debate between \citet{rosen1981} and \citet{adler1985} has stalled
because quality cannot be observed and success is a circular measure of it. The
design here does not observe quality either. It observes something the debate
has lacked: a detailed description of the content across images, speech and
text, recorded before outcomes are known and shown to carry real information.

The evidence favours Adler. Content predicts sub-genre but not returns, and
experience adds nothing; what changes a creator's position is a single hit that
cannot be predicted from anything measured about the content, and whose effect
then lasts. This is the pattern of path dependence described by
\citet{arthur1989} and \citet{david1985}, and it is what the experimental markets
of \citet{salganik2006} and \citet{salganik2008} produce under controlled
conditions. The same pattern appears here in a large real market, together
with a measure of how long the effect of a hit lasts.

Rosen's explanation survives only if talent lies in qualities that none of these
characteristics capture. A claim that winners are better must now say in what
way they are better, because the ways that industry practice points to do not
predict success.

\subsection{Implications for platforms}

The policy question this paper raises is what better feedback would be worth to
creators.

Creators already receive detailed statistics on their videos, and some can test
titles and thumbnails. None of it tells them which part of a hit paid, and
Section~\ref{sec:pays} shows the result: effort goes into copying a feature that
earns nothing, by creators who have no way to tell it apart from the feature that
does. Platforms are well placed to
solve this. They observe impressions and clicks for every video, and they can
run controlled experiments on what they display, which is exactly the
information needed to break a hit into its parts. Producers respond when a
platform changes what it rewards: after Facebook adjusted its news feed to
favour quality content from news publishers, German newspapers posted about
30\% more substantive political news on the platform \citep{garz2023}.

The gain from better feedback comes from redirecting effort creators already
spend rather than from asking them to work harder, and it is measurable: the
effort now spent copying the look of a hit could go into what actually pays.

\section{Conclusion}
\label{sec:conc}

In this market, nothing accumulates except a hit. Experience adds nothing, while
a single breakout raises the views
of everything published afterwards by 164\%. The hit cannot be predicted from
the content, even though the same content measures identify sub-genre with an
AUC of 0.782. When a hit arrives, creators copy it in two ways, and only one of
them pays.

These findings resolve two puzzles. Entering a market with a median return of
\$1.61 can be rational, because the prize is lasting rather than one-off, and
its value grows with every video a creator goes on to make. Copying the look of
a hit is not a failure of judgement either. A hit is a single event in which
every possible cause coincides, and neither the statistics nor the tests the
platform offers can separate them. Creators respond sensibly to the information
they have.

Better information could remove a real cost that this paper uncovers. One of
the two things creators copy after a hit, the look, has no measurable return.
Feedback that showed creators which part of a hit paid would redirect that
effort toward what does, without asking them to work any harder. Creators
already put in the effort; they lack the information to direct it.

The results hold under conservative assumptions. Under the strictest
definitions of a breakout, a hit still raises later views by more than 150\%.
After removing the trend before the breakout, the immediate jump is still 65\%.

Two directions for future research follow. First, impression-level data held by
platforms would separate the audience's response to a hit from the algorithm's.
Second, a cleaner test of what an image is worth is within reach. Creators often
replace the thumbnail on a video that is already published. This changes the
image while holding everything else fixed: the content, title, channel, topic,
audience, age and position in the catalogue. The change can also be measured in
exactly the characteristics studied here, such as a face added, text enlarged or
a palette warmed. The tools to collect such data are already built, and they
make this design the natural next step for research on what images are worth.

\section*{Funding}

This research received no funding from any source.

\end{document}